\documentstyle[twocolumn,aps]{revtex}

\newcommand{\beq}{\begin{equation}}
\newcommand{\eeq}{\end{equation}}
\newcommand{\beqa}{\begin{eqnarray}}
\newcommand{\eeqa}{\end{eqnarray}}
\newcommand{\beqar}{\begin{eqnarray*}}
\newcommand{\eeqar}{\end{eqnarray*}}

\def \la {\langle}
\def \ra {\rangle}
\def \up {\uparrow}
\def \down {\downarrow}

\input epsf
\begin{document}

\title{ How macroscopic properties dictate microscopic probabilities}

\author{    {\large Yakir Aharonov$^{a,b}$ and Benni Reznik$^{a}$} \\
 \small\it
a) School of Physics and Astronomy, Tel Aviv
                      University, Tel Aviv 69978, Israel.\\
b)  Department of Physics,
           University of South Carolina, Columbia, SC 29208.  }
\date{October 15, 2001}
\maketitle

\begin{abstract}
  We argue that the quantum probability law follows, in the large $N$
  limit, from the compatibility of quantum mechanics with
  classical-like properties of macroscopic objects. For a finite
  sample, we find that likely and unlikely measurement outcomes are
  associated with distinct interference effects in a sample
  weakly coupled to an environment.
\end{abstract}
\vspace {0.5 cm}

Given that quantum theory is probabilistic, is there a fundamental
physical principle that dictates the form of the quantum probability
law?  We will argue that, indeed, the assigned probabilities for a
given test performed on a large sample of identical systems, is
dictated by the classical-like properties of macroscopic systems.

Consider a sample of $N$ identically prepared non-entangled spins in
the state $|\psi\ra$.  The state of the sample is then
\beq
|\Psi\ra =|\psi\ra_1|\psi\ra_2 ....|\psi\ra_N . 
\eeq
Since spins carry a magnetic moment, the sample may also be viewed as
a collection of magnetic moments, all pointing in the same direction,
which under a suitable arrangement describe a magnet-like object. In the
limit of large enough $N$, the sample becomes a macroscopic object,
with definite collective properties like a total magnetic moment and
associated magnetic field.

Suppose now that we like to measure such a collective property of our
sample, say, the magnetic moment $M_x=\hat x\cdot \vec M$, in the
$\hat x$ direction. Since the magnetic moment of each spin is
proportional to the spin itself, we need to measure the $\hat x$ 
component of an observable like
\beq
\vec M = {\sum_{i=1}^N \vec \sigma_{i} \over N} . 
\eeq
In the large $N$ limit
\beq
\lim_{N\to \infty} [M_x, M_y] = i\lim_{N\to \infty} {M_z\over N} =0 .
\label{comm}
\eeq
This suggests that
averaged collective observably, like $\vec M$, represent
``macroscopic'', classical-like, properties of the sample.

Our main idea is to compare between two distinct methods, the
``macroscopic'' and ``microscopic'' methods, for observing the
macroscopic collective magnetic moment.  First, we consider a {\em collective}
measurement of $M_x$, which does not probe the state of individual
spins. For instance a single charged test particle can be scattered to
determine the total magnetic field of the sample. As we show below,
and as eq. (\ref{comm}) above indicates, in the large $N$ limit the
state is an eigenstate of $M_x$ and the outcome given by the
expectation value is deterministic .  The macroscopic measurement
induces a vanishing small disturbance to each individual spin.  Yet
the total accumulated effect on the scatterer is finite.  In the
second microscopic method, we measure separately on each spin the
operators $\sigma_{xi}$, $i=1...N$, and evaluate from the $N$ outcomes
the average of $M_x$. The microscopic measurements does disturb the
spins and randomizes the state of the sample according to the quantum
probability law.  However, since the first macroscopic method nearly
does not affect the sample its outcome should agree up to $1/\sqrt{N}$
corrections with the the mean result of a subsequent microscopic
measurement.  We shall see that in the limit of $N\to \infty$ this
suffices to fix the form of the quantum probability distribution.
On the other hand, for a large but finite 
sample, the probability law still follows if we make a further 
assumption on the stability of physical laws against small perturbations.

\noindent
Consider the following identity.  Operating with the spin operator
$\sigma\equiv \hat n \cdot \vec \sigma$, on a single spin state we can
express the resulting state as 
\beq \sigma |\psi\ra = \bar \sigma 
|\psi\ra + \Delta \sigma | \psi^\bot\ra ,
\label{identity}
\eeq
where  $\bar \sigma $  and $\Delta \sigma$, are the expectation value and
the uncertainty of $\sigma$ with respect to the state $\psi$,  and
$\psi^\bot$ is a normalized state orthogonal to $\psi$:
\beq
\la \psi |\psi^\bot\ra =0, \ \ \  ||\psi||=||\psi^\bot|| =1 .
\eeq
If $\sigma$ is not an eigenoperator of $|\psi\ra$ we have
on the right hand side of (\ref{identity}) also a component of $|\psi^\bot\ra$.
However let us now applying the relation to the collective
state $M_x|\Psi\ra$.  We get

\beqa
M_x|\Psi\ra &=& {\sum_{k=1}^N\bar \sigma_{xk} \over N}|\Psi\ra
+ {\Delta \sigma\over N} \sum_{k=1}^N |\Psi^{\bot}_k\ra \nonumber \\
&=& \bar\sigma_x |\Psi\ra + {\Delta \sigma\over N} |\Psi^\bot\ra .
\label{M}
\eeqa

Here $|\Psi^{\bot}_k\ra = |\psi\ra_1...|\psi^\bot\ra_k...|\psi\ra_N$.
Since $\la \Psi^\bot_i|\Psi_j^\bot\ra=\delta_{ij}$, the norm
of the second term on the right hand side is
\beq
||{\Delta \sigma\over N} |\Psi^\bot\ra || = {\Delta \sigma \over N} .
\eeq
Apart from the special cases that $\bar\sigma_x  \sim O(1/\sqrt{N})$),
the last term on the right hand side in (\ref{M})
is a small correction,
and in the large $N$ limit,
\beq
\lim_{N\to \infty} M_x|\Psi\ra = \bar \sigma_x  |\Psi\ra . 
\label{Mlimit}
\eeq
Similar results,
 which we discuss in the sequel, 
have be described in \cite{Hartle,Graham,FGG}.

Next, let us see that
the disturbance caused to individual spins, as a result of
a collective measurement of $M_x$, vanishes in the large $N$ limit.
The evolution of the system under a measurement
is described by the unitary operator $U=\exp (iQM_x)$, where
$Q$ is conjugate to the ``pointer'' $P$ of the measuring device.
Denoting by $|P\ra$ the initial state of the measuring device, (say
a Gaussian centered around $P$), and
applying $U$ to the combined state we have
\beq
U|\Psi\ra|P\ra = \bigotimes_{k=1}^{N}u_k|\psi_k\ra|P\ra ,
\eeq
where $u_k=e^{i{\sigma_{xk}\over N}Q}$.
Using eq. (\ref{identity}) we have 
\beq
u_k|\psi_k\ra = \big[ \cos{Q\over N} + i\bar\sigma_x \sin{Q\over N} \bigr]
|\psi_k\ra
  + i\Delta \sigma \sin{Q\over N}|\psi^\bot_k\ra .
\eeq
Expanding the later in $1/N$ we get
\beq 
\label{Ush}
U|\Psi\ra = 
\bigl[1-{\Delta\sigma^2Q^2\over2N}\bigr]e^{i\bar\sigma_x Q}|\Psi\ra|P\ra + 
 |\delta \chi\ra , 
\eeq
where
\beq
|\delta \chi\ra=  i {\Delta \sigma Q\over N} 
 \sum_{k=1}^N|\Psi^\bot_k\ra|P\ra + O\bigl({1/N^2}\bigr) . 
\label{deltachi}
\eeq
For non-zero $|\delta \chi\ra$ the measuring device is entangled with 
sample. 
Since $|\Psi^\bot_k\ra$ are mutually orthogonal, 
$\la \delta \chi|\delta \chi\ra \sim 1/N$, and the entanglement 
produced by this measurement is small.
Particularly, in the limit $N\to\infty$,
we may drop the term $|\delta\chi\ra$ above
and obtain
\beqa
\lim_{N\to \infty}U |\Psi\ra|P\ra &=& \exp( iQ \bar\sigma_x )|\Psi\ra|P\ra
\nonumber \\
&=& |\Psi\ra|(P-\bar\sigma_x) \ra .
\eeqa

For a given initial state, $|\psi\ra= c_+|+\ra + c_-|-\ra$
with $|+\ra$ and $|-\ra$, as the eigenstates of $\sigma_x$, 
the collective measurement will
shift the pointer by the value 
\beq
\bar\sigma_x =|c_+|^2(+1) + |c_-|^2(-1). 
\eeq

Next suppose that we 
perform on the {\em same} sample a ``microscopic''
measurement of each individual spin in the $\hat x$-direction. 
The outcome of this microscopic measurement is 
the string of numbers of $+1$ or $-1$.
The numbers,  $n_+$, of  $+1$'s, and
 $n_-=N-n_+$, of $-1$'s, should again allow us to evaluate
the average 
\beq
F_N= {n_+\over N}(+1) + {n_-\over N}(-1) .
\eeq
Since in this limit, the disturbance caused to the system vanishes, 
consistency of the microscopic measurements with
the macroscopic collective measurement
dictates the equality
\beq
\lim_{N\to \infty} F_N = \bar\sigma_x 
\eeq
or
\beq
\lim_{N\to \infty} \biggl[ {n_+\over N}(+1) + {n_-\over N}(-1)
\biggr] =  |c_+|^2(+1) + |c_-|^2(-1)
\label{prob}
\eeq
On the left hand side we have the averages obtained from the individual
measurements, while in the right hand side the deterministic result
for the macroscopic measurement of $\bar\sigma_x $.
From the above equation we identify $|c_{+}|^2$ and $|c_-|^2$
with the usual quantum mechanical frequencies for $\sigma_x=+1$ and
$\sigma_x=-1$.

Our argument can be easily generalized. For an n-level system
a single macroscopic measurement is not sufficient, because
the average is determined from the absolute square of $n$ amplitudes. 
However, now we can measure macroscopically 
$n-1$ independent commuting observables (e.g. $L_x, ..., L_x^{n-1}$)
and together with the overall normalization, $\sum n_i=N$,
evaluate the relevant probabilities.

The above result for the $N\to \infty$ limit is still unsatisfactory, 
when considering a finite sample, because in this case
we can no longer neglect the disturbance $|\delta\chi\ra$ in eq. 
(\ref{Ush}).
One way to proceed  is to 
make an additional assumption which seems natural
for macroscopically large samples:
{\em the results of physical
experiments are stable against small perturbations}.
Hence, for finite large $N$,
the operator $M_x$ fails to be a precise eigenoperator of $|\Psi\ra$
in eq. (7).
However,  by a small modification of the state to
$|\Psi\ra+|\delta\Psi\ra$,  with
magnitude $|| \  |\delta \Psi\ra|| = O(1/N)$, 
the perturbed state does become
an exact eigenstate of $M_x$. 
Alternatively, if after coupling to $P$, we first project 
the state in eq. (\ref{Ush}) by $|\Psi\ra\la \Psi|$ this 
would eliminate the term $|\delta \chi\ra$ and the final 
microscopic measurement
would give rise only to likely outcomes.
The ``stability'' principle, than requires that
the corresponding probability law can qualitatively change
at most by terms of order $O(1/\sqrt{N})$.

To get further insight to the physical meaning of this 
disturbance it is useful to adopt another
line of considerations.
Suppose that after coupling the sample to the pointer $P$, we 
do not observe the exact value of $P$, but proceed to perform the
 microscopic  measurement. Hence we now regard the pointer 
as an environment-like system that couples weakly with the sample.
The outcome of the microscopic measurement is 
described by the post-selected state $|n_+,n_-\ra$ of the sample. 
We can evaluate the final state of the pointer-system $P$ by projecting 
eq. (\ref{Ush}) from the left by $\la n_+,n_-|$:
\vfill\eject
$$
\la n_+, n_{-} | U |\Psi\ra|P\ra =
\bigl[1-{\Delta\sigma^2Q^2\over2N}\bigr]\times \ \ \ \ \ \  $$
\beq
\la n_+,n_- |e^{i\bar\sigma_x Q}\ra |\Psi\ra |P\ra 
 + \la n_+,n_-|\delta\chi\ra .
\eeq
Noting that $U$ is diagonal with respect to the final state of the sample
we get
\beq 
\label{md}
|P - F_N\ra = |P - \bar\sigma_x  \ra + |\delta P\ra
\eeq
where 
\beq 
\label{dp}
|\delta P\ra=
{\la n_+,n_-|\delta\chi\ra \over
 \la n_+,n_-| \Psi\ra}.
\eeq

The last equation states that the difference between a pointer state
shifted by the frequency and by the mean is given by the correction
$|\delta P\ra$.  Consider the case of a ``likely'' outcome, with $F_N-
\bar\sigma_x \approx 0 $.  By examining eqs.
(\ref{deltachi},\ref{md},\ref{dp}), we find that in this case
destructive interferences reduce the magnitude of the correction to
$||\delta P\ra ||\approx 0$.  On the other hand, consider now an
unlikely result $n_+=N, \ \ n_-=0$, hence $F_N- \bar\sigma_x \approx N
$. In this case the equality of the two sides in eq. (\ref{md}) cannot
be satisfied if $|\delta P\ra$ is small.  Indeed we get that in this
case, the post-selected environment state, interferes constructively
and $||\delta P\ra||\sim 1$. (In fact in this case all higher orders
in $1/N$ give rise to order one contributions.)  Hence what we could
have regarded in the likely case as negligible $1/N$ corrections
becomes now the dominant contribution.

It is interesting that here we see, as far as we know, for the first
time, a fundamental difference, from a {\em microscopic point of
  view}, between likely and unlikely outcomes for a given sample. The
unlikely results require a large coherent interference effect between
the microscopic amplitudes in (\ref{deltachi}), that are induced by the
weak interaction with the weakly coupled environment.

We further emphasis, that the analysis considered here is quite
general.  In reality, while a sample is measured, it is always
subjected to environmental effects which couples weakly with the
particles of the sample (e.g. the electromagnetic dipole coupling).

Before concluding, and for completeness, let us examine the effect of
a macroscopic measurement, from the point of view of the full quantum
formalism. (Hence from now on, we will assume the usual quantum
probability law.)  When we perform a precise measurement of $M_x$ we
must disturb other non-commuting observables.  In the case of the
individual microscopic measurements, we will randomly changed the
state of individual spins, and consequently destroy the macroscopic,
magnet-like, properties of the sample: the new state may look like a
collection of randomly polarized spins.

Now let us consider the collective measurement.  Clearly, when we
measure a macroscopic quantity (here the average magnetic moment of
the magnet) we do not destroy the macroscopic state.  However to be in
conformity with the uncertainty principle we must cause {\em some}
disturbance to the spins.  We will now show that the strength of the
disturbance is generally such that {\em not even one spin} of the
sample has flipped its direction. To do that we will now regard the
measuring device on a quantum level as well.

The accuracy of the collective measurement is determined by the
initial uncertainty of the measuring device pointer.
Let us express it as
\beq
\Delta P = {1\over \epsilon \sqrt N} ,
\eeq
where $\epsilon$ is some real number to be fixed in the sequel.
The disturbance to the $i$'th spin of the sample is then induced by the evolution
\beq
U_i = \exp i{Q \sigma_{xi}\over N} ,
\eeq
which describes a rotation around the $\hat x$ axis
of the spin, by an uncertain angle
of $\Delta \theta \approx {\Delta Q\over N} \approx {\epsilon \over \sqrt N}$.
The probability of a single spin to remain in its initial state (not to flip)
hence goes like $\approx
1- {\epsilon^2\over N}$.
Therefore the probability all the $N$ spins to
remain in their initial state is
\beq
\biggl(1-{\epsilon^2\over N}\biggr)^N \approx
\exp (-\epsilon^2) .
\eeq
Therefore, for $\epsilon\ll 1$, the probability that even one spin of the
sample has not flipped is close to unity.
Nevertheless, since at the same time, we can still have $N \gg 1/\epsilon^2$,
the measurement becomes arbitrarily precise for large enough $N$.

This confirms with the uncertainty relation for components of
$\vec M$ and a single spin. From the commutation
relation  (\ref{comm}) with a finite $N$, and from
 $ [\sigma_{zi}, M_x] = {i\sigma_{yi}\over N}$
we have
\beq
\Delta M_x \Delta M_y \ge {\la M_z\ra\over N}
\eeq
and
\beq
\Delta\sigma_i\Delta M(N) \ge {\la \sigma_{yi}\ra\over N}
\eeq

Hence we can measure simultaneously all components of $\vec M$
provided that we keep the accuracy as
$\Delta M_x \sim \Delta M_y \sim {1/\sqrt N}$.
Although, for large $N$ this inaccuracy becomes vanishing small,
we still  cannot distinguish between different eigenvalues
whose separations goes like $1/N$.

Finally, let us compare our approach with other related attempts to
derive quantum probabilities. Hartle and Graham \cite{Hartle,Graham}
have constructed the ``frequency operator'', $\hat f_N$ by considering
the sum of projectors to all possible results of the measurements,
that are weighted by an appropriate ``degeneracy'' factor. The latter
corresponding to the number of different strings of possible results
with the same total number of $n_+$ and $n_-$.  In the limit of $N\to
\infty$, $\hat f_{N}$ becomes an eigenoperator of $|\Psi\ra_{N}$.  The
eigenvalue of the frequency operator is then given by the quantum
probability. In our approach it is the collective magnetic operator
$\vec M$ that is becomes an eigenoperator.  However since in order to
observe $\hat f$, one has to perform a highly non-local measurement of
the spins sample, the frequency does not seem to be a physical
realizable operator.

Instead, we argued that in {\em ordinary}, everyday, macroscopic
observations we do measure collective operators like the operator
$\vec M$ discussed here.  Indeed, such collective operators, appears
naturally in usual electromagnetic interactions of an external test
particle with a sample of magnetic moments that constitute a
macroscopic object. An approach similar to ours has been suggested by
Farhi, Goldstone and Gutmann \cite{FGG}. The present
article extends this approach by analysing the measuring process.

In conclusion, we have demonstrated how the agreement between
macroscopic and microscopic observations dictates the quantum
probability law. For a finite sample we suggested that this law can be
obtained if one further invokes a natural stability assumption.

Finally, we compared between two definite measurement outcomes: a rare
sequence where all spins are found to be in the $\up$ direction, and an
expected sequence with roughly an equal number of $\up$ and
$\down$ outcomes.  Classically there is no microscopic difference
between the two sequences: they have the same a priori probability.
Surprisingly, we observed that in quantum mechanics these ``likely''
and ``unlikely'' sequences do differ on a microscopic level, and  are
associated with distinctive interference effects.

\vspace {1cm}

This research was supported in part by grant 62/01-1 of
the Israel Science Foundation, established by the
Israel Academy of Sciences and Humanities,
NSF grant PHY-9971005, and ONR  grant N00014-00-0383.

\end{document}